\documentclass[%
 reprint,
 aps,
]{revtex4-1}
\usepackage{amsmath}
\usepackage{amssymb}
\usepackage{upgreek}
\usepackage{graphicx}% Include figure files
\usepackage{dcolumn}% Align table columns on decimal point
\usepackage{bm}% bold math
\usepackage[mathlines,columnwise]{lineno}% Enable numbering of text and display math
\newcommand*\patchAmsMathEnvironmentForLineno[1]{%
  \expandafter\let\csname old#1\expandafter\endcsname\csname #1\endcsname
  \expandafter\let\csname oldend#1\expandafter\endcsname\csname end#1\endcsname
  \renewenvironment{#1}%
     {\linenomath\csname old#1\endcsname}%
     {\csname oldend#1\endcsname\endlinenomath}}% 
\newcommand*\patchBothAmsMathEnvironmentsForLineno[1]{%
  \patchAmsMathEnvironmentForLineno{#1}%
  \patchAmsMathEnvironmentForLineno{#1*}}%
\AtBeginDocument{%
\patchBothAmsMathEnvironmentsForLineno{equation}%
\patchBothAmsMathEnvironmentsForLineno{align}%
\patchBothAmsMathEnvironmentsForLineno{flalign}%
\patchBothAmsMathEnvironmentsForLineno{alignat}%
\patchBothAmsMathEnvironmentsForLineno{gather}%
\patchBothAmsMathEnvironmentsForLineno{multline}%
}

\usepackage{xcolor}

\begin{document}

\preprint{APS/123-QED}

\title{Noise-tunable nonlinearity in a dispersively coupled diffusion-resonator system using superconducting circuits}
%\title{Dispersively coupled diffusion-resonator systems using superconducting circuits}% Force line breaks with \\

\author{Christin Rh\'en}
\email{christin.rhen@chalmers.se}
\author{Andreas Isacsson}

\affiliation{Department of Physics\\
Chalmers University of Technology\\
SE-412 96 G\"oteborg\\
Sweden}

\date{\today}% It is always \today, today,

\begin{abstract}
The harmonic oscillator is one of the most widely used model systems in physics: an indispensable theoretical tool in a variety of fields. It is well known that otherwise linear oscillators can attain novel and nonlinear features through interaction with another dynamical system. We investigate such an interacting system: a superconducting LC-circuit dispersively coupled to a superconducting quantum interference device (SQUID). We find that the SQUID phase behaves as a classical two-level system, whose two states correspond to one linear and one nonlinear regime for the LC-resonator. As a result, the circuit's response to forcing can become multistable. The strength of the nonlinearity is tuned by the level of noise in the system, and increases with decreasing noise. This tunable nonlinearity could potentially find application in the field of sensitive detection, whereas increased understanding of the classical harmonic oscillator is relevant for studies of the quantum-to-classical crossover of Jaynes-Cummings systems.
\end{abstract}

\maketitle

The harmonic oscillator is one of the most well-understood dynamical systems in physics, and is used as a model in nearly every field. The classical harmonic oscillator was studied already by  Galileo Galilei, while its quantum counterpart was described in 1925 by Paul Dirac. It remains one of few models that can be exactly solved, and as such, it features prominently in courses on classical and quantum mechanics. It is perhaps surprising, then, that the harmonic oscillator still remains at the forefront of contemporary physics research.

Today, considerable attention is devoted to harmonic oscillators that interact with an auxiliary dynamical system. This situation appears, for instance, in circuit quantum electrodynamics~\cite{Abdo_2009,Fink_2010,Yamamoto_2014} and quantum information processing~\cite{Chiorescu_2004,Lupascu_2004,Irish_2005,Schuster_2007,Petersson_2012}, where the harmonic oscillator models a superconducting microwave circuit and the auxiliary system is a qubit. Then, manipulation of the circuit allows for control and read-out of the state of the qubit. In a similar manner, when the auxiliary system is a second harmonic oscillator, as in optomechanics~\cite{Johansson_2014,Regal_2008,Thompson_2008,Palomaki_2013}, one of the oscillators can be damped or driven by manipulating the other.  

An additional interesting case is when another very common model system takes the role of auxiliary system: the diffusing Brownian particle. One proposed realization of such a coupled system is a diffusing particle loosely adsorbed on the surface of a nanomechanical resonator~\cite{Atalaya_2011a, Atalaya_2011b, Atalaya_2012, Edblom_2014, Rhen_2016}. Then, the particle position directly influences the oscillator's natural frequency, and the oscillator in turn provides an amplitude-dependent inertial back-action force on the particle. Despite its apparent simplicity, this diffusion-resonator system exhibits surprising effects such as induced nonlinearity~\cite{Atalaya_2011b} and bistability~\cite{Atalaya_2011a}, inhomogeneous dephasing~\cite{Atalaya_2012}, as well as mode coupling and non-linear dissipation~\cite{Edblom_2014, Rhen_2016}
. As shown recently~\cite{Dykman_2014}, these features are rather generic for a harmonic oscillator mode coupled dispersively to an auxiliary dynamical system, under certain circumstances. However, with the current state of the art, it is very difficult to fabricate this nanomechanical system in a parameter regime where an interesting physical response will be observable.

\begin{figure}[t]
    \centering
    \includegraphics[width=0.9\linewidth]{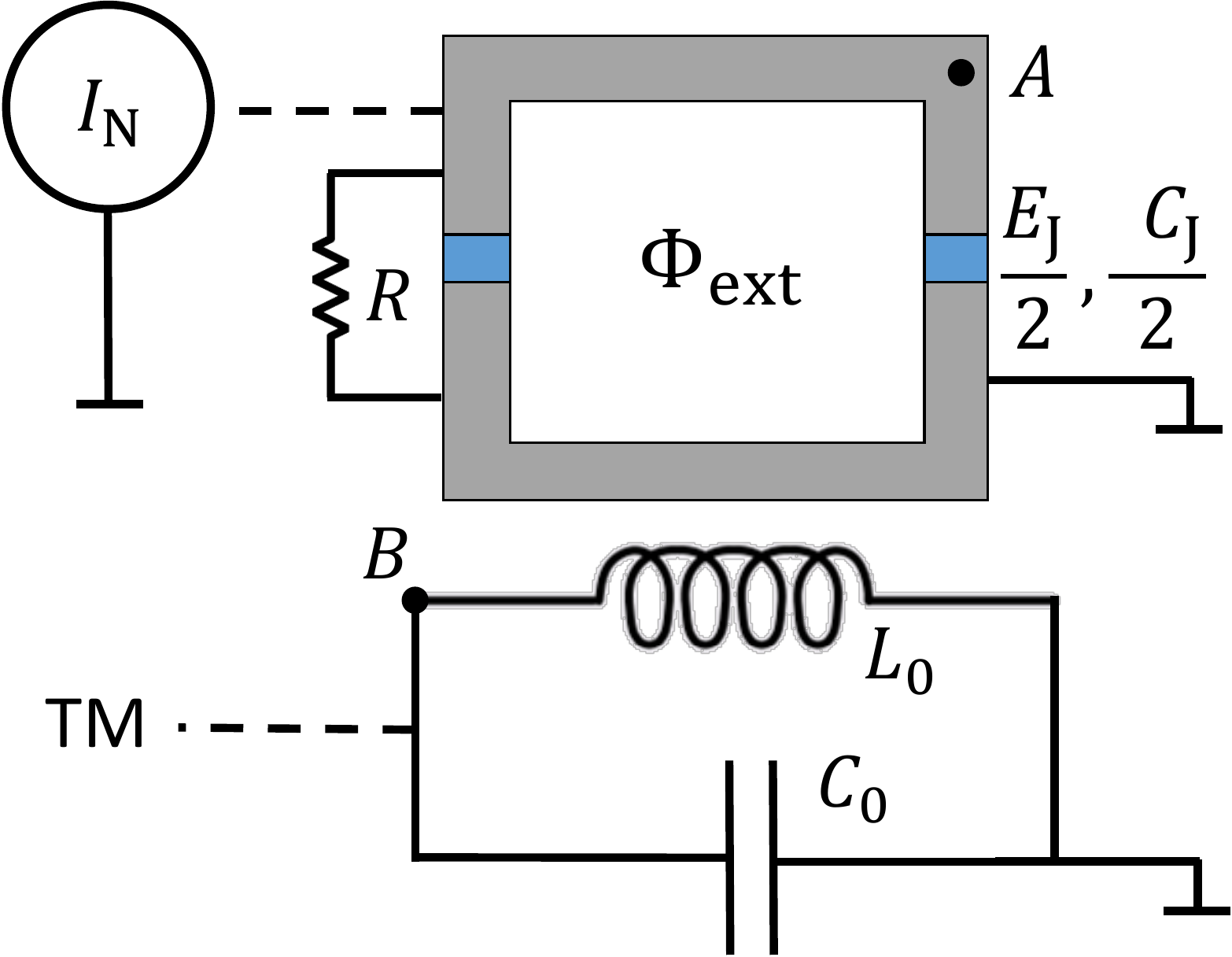}
    \caption{Superconducting circuit realization of a diffusion-resonator system. An overdamped symmetric SQUID is inductively coupled to a lumped LC-resonator. A weak inductive coupling ensures a dispersive interaction between the two systems. Additional noise can be introduced in the system using an external noise current source $I_{\rm N}$. For read-out and actuation, the LC-resonator can be coupled inductively or capacitively to an external transmission line. The nodes $A$ and $B$ indicate where the node fluxes used below as dynamic variables are defined.}
    \label{fig:system}
\end{figure}

 Here we propose an alternative realization of a resonator-diffusion system, making use of superconducting circuit elements. These allow for a high degree of control over the relevant parameters, some of which can be tuned in situ. Our proposed realization, depicted in Fig.~\ref{fig:system}, makes use of a resistively shunted superconducting quantum interference device (SQUID), whose phase variable will act as a diffusing particle, due to the presence of noise in the shunting resistor. The harmonic oscillator is represented by a lumped superconducting LC-resonator, and is inductively coupled to the SQUID. We find that when the resonator is driven, the SQUID phase locks in to one of two values; it behaves as a classical two-level system. Interestingly, the LC-circuit exhibits dramatically different dynamics for the two values of the phase. In one case the circuit becomes a linear oscillator, while in the other it is highly nonlinear and can be multistable. As the resonator amplitude increases, the system switches between linear and non-linear regimes in a quasi-periodic manner, determined by the noise level and the drive amplitude. We derive an analytical model that is very successful at predicting the two regimes, and discuss where this model breaks down. 

While it is clear that the tunable nonlinearity found in the studied circuit could find application in the field of sensitive detection (c.f. Josephson bifurcation amplifiers~\cite{Vijay_2009}), our results also have more fundamental implications. In the quantum regime, a two-level system coupled to a harmonic oscillator is described by the well-studied Jaynes-Cummings Hamiltonian. However, an understanding of the transition between this quantum regime and its classical counterpart remains elusive. As we here investigate the \emph{classical} dynamics of a harmonic oscillator coupled to a two-level system, new light is shed on the less-known half of this quantum-to-classical crossover.

\section{Results}
\subsection{Circuit description}
We study a system consisting of a symmetric SQUID in the vicinity of a lumped LC-resonator with capacitance $C_0$ and inductance $L_0$, as shown in Fig.~\ref{fig:system}. For actuation and readout purposes, the resonator can be coupled to an external transmission line. Each Josephson junction in the SQUID is characterized by a capacitance $C_{\rm J}/2$ and Josephson energy $E_{\rm J}/2$. We take as dynamic variables the fluxes $\Phi_{\rm A,B}\equiv\int_{-\infty}^t{\rm d}t'\,V_{\rm A,B}(t')$ at points $A$ and $B$, respectively (see Fig.~\ref{fig:system}). The conservative dynamics of the system is then described by the Lagrangian 
\begin{align}
L&=\frac{1}{2}C_0\dot{\Phi}_{\rm B}^2-\frac{1}{2}\frac{\Phi_{\rm B}^2}{L_0}+\frac{1}{2}C_{\rm J}\dot{\Phi}_{\rm A}^2\nonumber\\
&\quad-E_{\rm J}\cos \frac{\pi\Phi_{\rm ext}}{\Phi_0}\left(1-\cos\frac{2\pi\Phi_{\rm A}}{\Phi_0}\right), 
\label{eq:lagrange}
\end{align}
where $\Phi_0=h/2e$ is the flux quantum. The external flux $\Phi_{\rm ext}$ threading the SQUID is partly determined by the current in the LC-circuit. If there are no other sources of external flux, the coupling is linear to lowest order in $\Phi_B$; $\Phi_{\rm ext}\approx 2g\Phi_B$.

Defining nondimensional variables $q=2\pi g\Phi_{\rm B}/\Phi_0$ and $x=\Phi_{\rm A}/\Phi_0$, Eq.~(\ref{eq:lagrange}) leads to the equations of motion 
\begin{gather}
\ddot{q}+\frac1{L_0C_0}q-(2\pi g)^2 \frac{ E_{\rm J}}{C_0\Phi_0^2}\sin q\left(1-\cos2\pi x\right)=0,\\
\ddot{x}+(2\pi)^2\frac{ E_{\rm J}}{C_{\rm J}\Phi_0^2}\cos q\sin2\pi x=0.
\end{gather}
The SQUID has a normal total resistance $R$, arising from either junction-internal resistance or from additional shunting. That is, the equation for $x$ should be supplemented by damping and noise, leading to the equation for the resistively and capacitively shunted Josephson junction (RCSJ) model~\cite{Stewart_1968, McCumber_1968}:
\begin{equation}
\ddot{x}+\frac1{RC_{\rm J}}\dot{x}+(2\pi)^2\frac{E_{\rm J}}{C_{\rm J}\Phi_0^2}\sin\left(2\pi x\right)\cos\left(q\right)=\frac{\sqrt{D}}{RC_{\rm J}}\xi(t).
\end{equation}
The RCSJ-model has reliably been able to reproduce experimental results in regimes where $\hbar\omega_{\rm J}<k_{\rm B}T$~\cite{Blackburn_2016}, with $\omega_{\rm J}=2\pi\Phi_0^{-1}\sqrt{E_{\rm J}/C_{\rm J}}$ being the plasma frequency.

As noted already by Ambegaokar and Halperin~\cite{Ambegaokar_1969}, the SQUID phase dynamics is that of a particle executing Brownian motion in a potential. For purely thermal fluctuations, the current noise in the resistive component is given by the Callen-Welton formula $S_{II}(\omega)=({\hbar\omega}/{\pi R})\coth(\hbar\omega/2k_{\rm B}T)$. We consider here the classical limit $\hbar\rightarrow 0$, which allows us to treat the noise as Gaussian white noise $\left<\xi(t)\xi(t')\right>=\delta(t-t')$ with a diffusion constant given by $D=2k_BTR/\Phi_0^2$. However, it is also possible to impose noise externally by attaching a noisy current source $I_{\rm N}$, as shown in Fig.~\ref{fig:system}. 

Rescaling the time variable to dimensionless time \mbox{$\tau=\omega_0t$}, with $\omega_0=1/\sqrt{L_0C_0}$, the equations of motion reduce to the form
\begin{gather}
\ddot{q}+\gamma\dot{q}+q-\epsilon\sin q\left(1-\cos2\pi x\right)=f(\tau)\label{eq:qeq}\\
\ddot{x}+\eta\dot{x}+\alpha\cos q \sin 2\pi x=\eta\sqrt{\cal D}\xi(\tau)\label{eq:xeq}.
\end{gather}
Here, we have introduced a finite quality (Q-) factor $1/\gamma$ to the LC-resonator, and added the external drive $f(\tau)$. 

The nondimensional constants $\epsilon$, $\alpha$, $\eta$, and ${\cal D}$ entering Eqs.~(\ref{eq:qeq}) and~(\ref{eq:xeq}) are related to physical quantities through
\begin{align}
\epsilon&\equiv  \frac{(2\pi g)^2 E_{\rm J}}{C_0\Phi_0^2\omega_0^2}= \frac{2\pi g^2I_{\rm c}}{C_0\Phi_0\omega_0^2}=2\pi g^2\frac{I_{\rm c}}{I_0},\\
\alpha&\equiv \frac{(2\pi)^2 E_{\rm J}}{C_{\rm J}\Phi_0^2\omega_0^2}=\frac{2\pi I_{\rm c}}{C_{\rm J}\Phi_0\omega_0^2}=2\pi\frac{C_0}{C_{\rm J}}\frac{I_{\rm c}}{I_0},\\
\eta&\equiv \frac1{C_{\rm J}R\omega_0},\\
{\cal D}&\equiv\frac{2k_{\rm B}TR}{\omega_0\Phi_0^2}=\frac{2k_{\rm B}T}{I_0\Phi_0}RC_0\omega_0=\frac{T}{T_0}.
\end{align}
Here, $I_0\equiv C_0\Phi_0\omega_0^2$ is the characteristic scale for the current through the LC-circuit, and $I_{\rm c}=2\pi E_{\rm J}/\Phi_0$ is the critical current of the junctions. The temperature scale is $T_0=\Phi_0I_0/2k_{\rm B}RC_0\omega_0$. 

It should be noted that as the dissipative term $\gamma\dot q$ is introduced in Eq.~\eqref{eq:qeq}, the fluctuation-dissipation theorem dictates that a stochastic force component $f_{\rm N}=\gamma\sqrt{\mathcal D_q}\xi_q(\tau)$ is included in the external force $f$. The additive noise $f_{\rm N}$ is proportional to $\gamma=1/Q$, which, in LC-circuits at cryogenic temperatures, can easily be as small as $10^{-5}-10^{-6}$. Even for the comparably bad LC-resonator (with $\gamma = 10^{-3}$) used in our simulations, the effect of the flux-noise term in Eq.~\eqref{eq:xeq} is an order of magnitude larger than that from $f_{\rm N}$ with the parameters used here. %the term that would have appeared in Eq.~\eqref{eq:qeq} (in the parameter range discussed below). 
%\AI{I removed a bit of this discussion as it seemed like we were making excuses. We can put them back if the referee starts making a fuzz.}
%Furthermore, in our primary concern is nonlinear response in the  is primarily the driven regime that is of interest,  in which the driving force $f(\tau)$ will dominate any stochastic force. Hence, we conclude that the primary mechanism for noise-induced dynamics is fluctuations in the SQUID flux, and thermal fluctuations in the resonator flux can be ignored.  If instead the ringdown to a thermalized state was to be studied, the stochastic force would have to be included in Eq.~\eqref{eq:qeq}.

\subsubsection{Parameter values}
We consider a typical LC-circuit with  $L_0=1$~nH and $C_0=0.1$~pF, corresponding to an LC-frequency of $\omega_0=(L_0C_0)^{-1/2}=10^{11}$~s$^{-1}$. The resulting characteristic current scale for the LC-circuit is $I_0=2$~$\upmu$A. We further consider junctions with Josephson energy $E_{\rm J}=0.6$~meV and capacitance $C_{\rm}=1$~pF, that are shunted by a resistance $R=16\ \Omega$. Using an external shunt resistance, $I_{\rm c}$ becomes largely independent of $R$, and can be tuned by the temperature; here, we have $I_c=0.5$~$\upmu$A at a temperature $T=3$ K. These parameter values, along with the resulting dimensionless parameters, are listed in Table.~\ref{tab:params}.

For the nondimensional coupling constant $g$ between a SQUID and the resonator we use $g=0.1$, and set the LC-resonator inverse Q-factor to $\gamma=0.001$. The relatively large values of $g$ and $\gamma$ were chosen for computational convenience. As will be seen from the analysis in sec.~\ref{sec:analytics}, for weaker coupling $g$ the same response will occur, provided the resonator damping $\gamma$ is reduced accordingly. 

\renewcommand{\arraystretch}{1.25}
\begin{table}[t] 
\caption{Typical values for circuit parameters, used in the simulations below. The corresponding dimensionless parameters are listed in the right column. The simulation temperature $T$ was chosen as 3 K, and the Q-factor of the LC-resonator was set to 1000. }

    \centering
    \begin{tabular}{cccc|cccc|cccc}
    \hline
     & $T_0$      & 1000 K     & & & $E_{\rm J}$      & 0.6 meV      & & & $\gamma$   & 0.001  & \\
     & $L_0$      &  1 nH      & & & $R$              & 16 $\Omega$  & & & $\epsilon$ & 0.015  & \\
     & $C_0$      & 0.1 pF     & & & $C_{\rm J}$      & 1 pF         & & & $\alpha$   & 0.15  & \\
     & $\omega_0$ & 100 GHz    & & & $\omega_{\rm J}$ & 3 GHz        & & & $\eta$     & 0.625   & \\
     & $I_0$      & 2 $\upmu$A & & & $I_{\rm c}$      & 0.5 $\upmu$A & & & ${\cal D}$ & 0.003 & \\
    \hline
    \end{tabular}
    \label{tab:params}
\end{table}

We will consider a periodic driving force $f(\tau)=f_0\cos\Omega\tau$, where $f_0$ is dimensionless. In practice, this coefficient is related to the number of drive photons. The mean photon occupancy of the LC-resonator is 
\begin{equation}
n_{\rm ph}=\frac{\langle E_{\rm LC}\rangle}{\hbar\omega_0}=\frac{\langle q^2\rangle\Phi_0^2}{8\pi^2g^2\hbar}\sqrt{\frac{C_0}{L_0}}=\frac{f_0^2\Phi_0^2}{16\pi^2g^2\gamma^2\hbar}\sqrt{\frac{C_0}{L_0}},
\end{equation}
where the resonator energy $E_{\rm LC}=\Phi_B^2/2L_0$ and $\langle q^2\rangle = f_0^2/2\gamma^2$ for an unperturbed resonator ($\epsilon=0$) driven at resonance. Substitution of the parameter values of Table~\ref{tab:params} yields $n_{\rm ph}\approx 10f_0^2/\gamma^2$. For an inverse Q-factor $\gamma=10^{-3}$ and drive amplitudes $f_0\simeq 10^{-2}$, we thus find $n_{\rm ph}\approx 10^3$.

\subsubsection{Equivalence with nanomechanical system} %consider subsection title
\begin{figure}[t]
    \centering
    \includegraphics[width=0.9\linewidth]{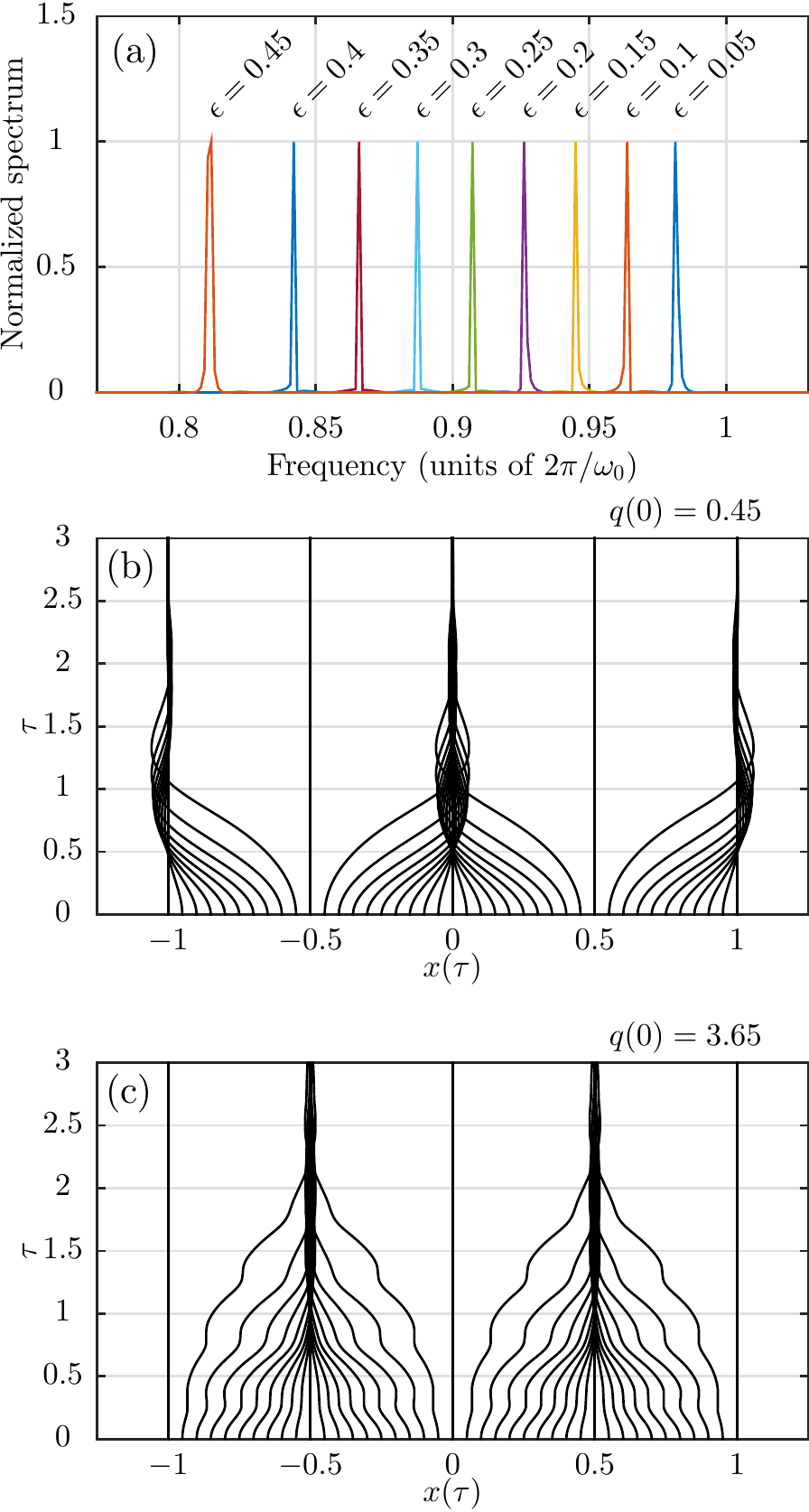}
    \caption{Numerical integration of the full equations of motion~\eqref{eq:qeq}-\eqref{eq:xeq}, absent external drive and noise ($f(\tau)=0$, $\gamma=0$, $\eta = 0$, and $\mathcal D=0$). (a) Resonator power spectra for increasing $\epsilon$. The phase of the SQUID is rapidly trapped at (b) integer or (c) half-integer values of $x$, depending on resonator amplitude. Note that the quite large values of $\epsilon$ used in (a) are chosen to illustrate the frequency shift clearly; the $\epsilon=0.01$ used in (b)--(c) would correspond to a much smaller frequency shift.}
    \label{fig:overdamped}
\end{figure}

To connect Eqs.~\eqref{eq:qeq}-\eqref{eq:xeq} to the nanomechanical resonator-physical particle system studied in Refs.~\cite{Atalaya_2011a, Atalaya_2011b, Atalaya_2012, Edblom_2014, Rhen_2016}, we note that for a small amplitude $|q|\ll 1$, Equations~(\ref{eq:qeq}) and~(\ref{eq:xeq}) resemble the equations of motion of a particle diffusing on a vibrating string. Expanding the trigonometric terms and identifying the vibrational mode function $\varphi(x)=\sqrt 2\sin\pi x$, we find
\begin{gather}
\ddot{q}+\gamma\dot{q}+[1-\epsilon\varphi^2(x)]q=f(\tau)\label{eq:qeq2}\\
\ddot x + \eta\dot{x}+\alpha\partial_x[\varphi^2(x)](1-q^2/2)=\eta\sqrt{\cal D}\xi(\tau)\label{eq:xeq2}.
\end{gather}
This is exactly the single-mode equations of motions seen in Ref.~\cite{Edblom_2014}, with the addition that the unperturbed \mbox{($q=0$)} motion of the particle described by $x$ is no longer free diffusion. Instead, the unperturbed SQUID phase moves in a spatial potential $\mathcal V(x)\propto \varphi^2(x)$, whose minima are at integer values of $x$. The presence of the "particle" causes the frequency of the LC-resonator to shift downwards, as shown in Fig.~\ref{fig:overdamped}~(a), while the $q$-dependent ``inertial'' force %circuit analogue? electrostatic potential?
drives the particle towards an antinode of $\varphi(x)$, seen in Fig.~\ref{fig:overdamped}~(b). The nanomechanical and the superconducting circuit systems thus show essentially the same dynamics, when taking into consideration the potential $\mathcal V(x)$. 

\subsubsection{Linear and nonlinear regimes}
In the superconducting circuit, the total effective potential that determines the dynamics of the SQUID flux $x$ is a combination of the effective potential created by the oscillation in the LC-circuit, that traps the flux near \mbox{$x_{\rm eq.}=n+1/2$}, $n\in\mathbb Z$, and the potential $\mathcal V(x)\propto\sin^2\pi x$, that traps the flux near $x_{\rm eq.}=n$. The steady-state value $x_{\rm eq.}$ thus depends on the relative strength of these two effects, which is determined by the resonator amplitude; see \mbox{Fig.~\ref{fig:overdamped} (b)-(c).} 

This interaction of two periodic potentials causes the resonator-SQUID system to display very interesting action-backaction dynamics. The resonator amplitude determines the equilibrium position $x_{\rm eq.}$ of the flux particle. Both integer and half-integer $x_{\rm eq.}$ have in common that the supercurrent through the SQUID, $I_s=I_c\cos q\sin 2\pi x$, vanishes, and Eq.~\eqref{eq:xeq} reverts to the familiar Langevin equation. However, the value of $x$ has a dramatic impact on the dynamics of the resonator, tuning it from linear to highly nonlinear depending on whether $x_{\rm eq.}$ is integer or half-integer. 

When $x$ is an integer, the term proportional to $\epsilon$ in Eq.~\eqref{eq:qeq} vanishes, and the equation for the LC-resonator becomes that of a driven, damped oscillator. As such, it should exhibit a Lorentzian frequency response with maximum at $f_0/\gamma$ and width $\gamma$. When $x$ is a half-integer, the absolute value of the $\epsilon$-term is maximized; this is the maximally non-linear regime. For $f_0=0$ and $x = n+1/2$, Eq.~\eqref{eq:qeq} becomes $\ddot q+\gamma\dot{q}+q=2\epsilon \sin q$, which describes an inverted physical pendulum of the kind used in the Holweck-Lejay gravimeter~\cite{Coullet_2009}.

\subsection{Driven response\label{sec:analytics}}
We now turn to the driven response by considering a periodic driving force $f(\tau)=f_0\cos\Omega\tau$. The resonator amplitude will depend on the drive amplitude and frequency, and the value of $x$ will in turn depend on the resonator oscillation. 

First, we analytically estimate the system's response to the drive by analyzing it in the adiabatic, mean-field, rotating wave approximation. The full stochastic equations of motion are then numerically solved. Except for a small region of anomalous response, the agreement between the analytical and numerical solutions is excellent.

\subsubsection{Adiabatic RWA solution}
\label{sec:rwa}
To find the steady-state solution of the slow-moving envelope $|u|$ of the resonator oscillation, we make the change of variables $2q=\left(ue^{i\Omega\tau}+u^*e^{-i\Omega\tau}\right)$, $2\dot{q}={i\Omega}\left(ue^{i\Omega\tau}-u^*e^{-i\Omega\tau}\right)$. In the rotating wave approximation (RWA), the equations of motion~\eqref{eq:qeq}-\eqref{eq:xeq} transform into
\begin{gather}
    \dot u=\left[-\frac \gamma2 +i\left(\frac{\epsilon\varphi^2(x)}{|u|} J_1(|u|)-\sigma\right)\right]u+\frac{f_0}{2i},\label{eq:RWAu}\\
    \ddot{x}+\eta\dot{x}+\frac\alpha{2\pi} J_0(|u|) \partial_x\left[\varphi^2(x)\right]=\eta\sqrt{\cal D}\xi(\tau).\label{eq:RWAx}
\end{gather}
Here, we have assumed that the detuning $\sigma=\Omega-1$ is small ($|\sigma|\ll1$), and we use $\varphi(x)=\sqrt 2\sin\pi x$ for brevity. Additionally, $J_{0,1}$ are Bessel functions of the first kind. Note that the coupling constant $g$ only affects $\epsilon$ in Eq.~(\ref{eq:RWAu}), and that a change in $\epsilon$ can be compensated for by a corresponding change in damping $\gamma$.

In the adiabatic limit, in which the relaxation time of the SQUID flux dynamics is much shorter than the relaxation time of the resonator ($\gamma\ll\eta$), the system state can be approximately described by a quasi-stationary probability distribution $p_{\rm st.}(x,\dot x,|u(t)|)$.
This distribution is derived by solving the Fokker-Plack equation corresponding to Eq.~\eqref{eq:RWAx}, under the assumption that $|u|$ is constant. The result is
\begin{equation}
    p_{\rm st.}(x,\dot x,|u(t)|)=\frac 1{\cal Z}\exp\left[-\frac{\dot x^2}{\eta\mathcal D}-\frac{\alpha}{\pi\eta\mathcal D}J_0(|u|)\varphi^2(x)\right],
\end{equation}
where ${\cal Z}=\sqrt{\eta\mathcal D\pi}\exp\left[-\frac\alpha{\pi\eta\mathcal D}J_0(|u|)\right]I_0\left(\frac{\alpha }{\pi\eta\mathcal D}J_0(|u|)\right)$.
Solving now for the stationary solution of Eq.~\eqref{eq:RWAu}, and making the mean-field approximation $\phi^2 \rightarrow \left<\phi^2\right>$, we arrive at the equation for the stationary amplitude $|u|$
\begin{equation}
|u|^2\left[\frac{\gamma^2}{4}+\left[\sigma-\epsilon h(|u|)\right]^2\right]=\frac{f_0^2}{4}\label{eq:ampleq}.
\end{equation}
The frequency shift $h(|u|)$ is given by
\begin{equation}
    h(|u|) =\frac{J_1(|u|)}{|u|}\left[\frac{I_1\left(J_0(|u|)\alpha/\pi\eta\mathcal D\right)}{I_0\left(J_0(|u|)\alpha/\pi\eta\mathcal D\right)}-1\right],
\end{equation}
where, $I_{0,1}$ are modified Bessel functions of the first kind. 

\begin{figure}[t]
    \centering
    \includegraphics[width=\linewidth]{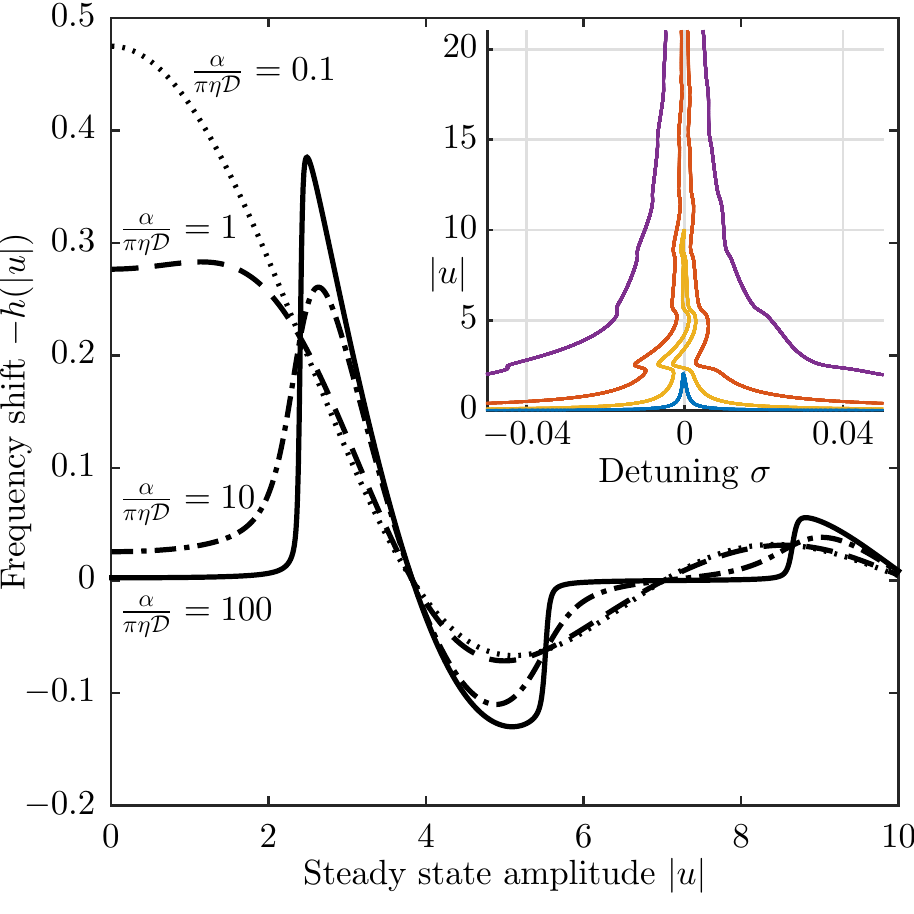}
    \caption{Frequency shift $h(|u|)$ as function of amplitude $|u|$ in the adiabatic, mean-field, rotating wave approximation for different values of the ratio $\alpha/\pi\eta{\cal D}$. In the low noise limit (solid curve), for small $|u|$ the phase particle is trapped at integer $x$, resulting in a zero frequency shift. For larger diffusion, the resonant frequency at low amplitudes increases with increasing noise ($\cal D$). The inset shows the corresponding frequency response curves as a function of detuning $\sigma = \Omega-1$, obtained by solving Eq.~(\ref{eq:ampleq}). Here, $\alpha/\pi\eta{\cal D}=25$, as given by the parameter values in Table~\ref{tab:params}, and we consider drive amplitudes $f_0=0.002$ (blue curve), $f_0=0.01$ (yellow curve), $f_0=0.04$ (red curve), and $f_0=0.2$ (purple curve). For moderate drive amplitudes $f_0\lesssim\epsilon$, multistability beyond bistability is possible.}
    \label{fig:gfun}
\end{figure}

Interestingly, only the ratio $\alpha/\pi\eta{\cal D}=I_{\rm c}\Phi_0/k_{\rm B}T=2I_c\Phi_0/RS_0$ (where  $S_{II}(\omega)=S_0$ for white noise) enters into the expression for the scaled frequency shift $\sigma$. This ratio can also be written as $2\pi E_{\rm J}/k_{\rm B}T$. In other words, the frequency shift, and hence the qualitative dynamics of the system, is determined by the ratio between the Josephson energy and the thermal energy.

In Fig.~\ref{fig:gfun}, the frequency shift $h(|u|)$ is shown for four different values of this ratio. For low resonator amplitudes (small $|u|$), the resonant frequency shifts downwards upon increasing the noise. As amplitude increases, either softening or hardening is observed depending on the noise power. The function $h$ has an infinite number of crossings with the horizontal axis, tending to zero as $|u|^{-3/2}$ in the limit of large $|u|$. 

The overall shape of the resonance curve, shown in the inset of Fig.~\ref{fig:gfun}, can be understood from treating the limits of high and low noise. In the low noise limit, ${\cal D}\rightarrow 0$, the `particle' coordinate $x$ will localize at integer values $x=n$ if $J_0(|u|)>0$ and at half-integer $x=n+1/2$ values if $J_0(|u|)<0$. The SQUID phase thus behaves as a classical two-level-system (TLS) whose state depends on amplitude of the resonator. In this limit, the frequency shift $h(|u|)$ approaches $2\left[\theta(J_0(|u|))-1\right]J_1(|u|)/|u|.$ Hence, the zeros of $J_0$ separates regions where the oscillator response is shifted away from or coincides with the unperturbed Lorentzian line shape: i.e., between regions where it has ordinary linear behavior and where it behaves as a driven Holweck-Lejay-like resonator. 

As noise increases, one sees from Fig.~\ref{fig:gfun} that the sharp features of the frequency shift $h(|u|)$ are smoothed out, and only at discrete amplitudes $|u|$ corresponding to zeros of $J_1$, will the frequency shift vanish. Hence, in the presence of noise the piecewise linear behavior in $|u|$ obtained for ${\cal D}=0$ is destroyed, and for strong noise $h(|u|)\rightarrow -J_1(|u|)/|u|$. We conclude that by varying the noise intensity, the frequency response can be tuned.

The shape of the response curve is also influenced by the drive strength. As expected, and also shown in the inset of Fig.~\ref{fig:gfun}, the resonance peak is Lorentzian for small $f_0$, but quickly takes on a flame-like character as the driving force increases. However, as the drive amplitude increases further, the response once more resembles a Lorentzian. This can again be traced back to the structure of the frequency detuning function $h(|u|)$, which decays algebraically with $|u|$. Consequently, the frequency shift near the top of the resonance peak quickly decays with increasing $f_0$. At the base of the resonance, on the other hand, the width of the peak is of order $f_0/|u|$, whereas the frequency detuning scales with $\epsilon$. Hence, we expect no visible nonlinear response when $f_0\gtrsim \epsilon$.

% It can be shown that order to observe at least bistability, the condition $\epsilon|u|^2\partial h/\partial|u|=\pm f_0^2/(2\sqrt{f_0^2-|u|^2\gamma^2})$ must be fulfilled for some $|u|$ on the resonance curve.

%\AI{  which decays as $|u|^{-3/2}$ for large $|u|$. Hence,  causing the induced nonlinearity to vanish for large amplitudes near the peak resonance. At the base of the resonance, the width of the resonance is of order $f_0/|u|$, whereas the frequency detuning scales as $~\epsilon$. Hence, we expect no visible nonlinear response when $f_0\gtrsim \epsilon$.  }

\subsubsection{Frequency response}
The stochastic equations of motion~\eqref{eq:qeq}-\eqref{eq:xeq} were numerically integrated using a second-order algorithm~\cite{Mannella_1989,Mannella_2000}, with the parameter values listed in Table~\ref{tab:params}. These correspond to $\alpha/\pi\eta{\cal D}=25$.

\begin{figure}[t]
    \centering
    \includegraphics[width=\linewidth]{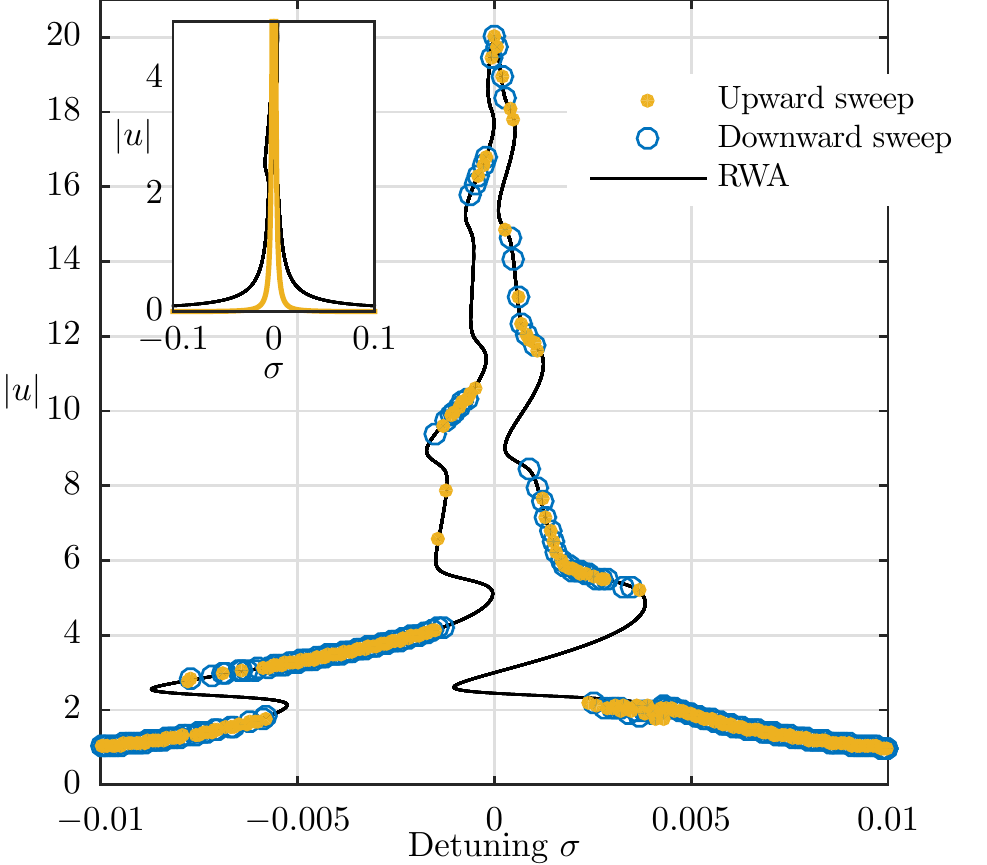}
    \caption{Simulated resonant response of the circuit, as the drive frequency is swept up (yellow dots) and down (blue circles). The black curve is the analytical response; the agreement is excellent. Here, the drive amplitude $f_0=0.02$, and $\alpha/\pi\eta{\cal D}=25$. Since temperature is finite, there is noise-induced switching between multistable states, and hysteresis loops are smeared. The inset shows the bottom of the resonant peak (black) together with the Lorentzian response of an unperturbed system (yellow) -- broadening is significant.}
    \label{fig:resonance}
\end{figure}

To begin with, the resonant response of the LC-circuit was calculated. The drive frequency $\Omega$ was varied while $f_0=0.02$, and the corresponding amplitude response found; the results are shown in Fig.~\ref{fig:resonance}. The agreement between simulation and the analytical results of Sec.~\ref{sec:analytics} is excellent.

\begin{figure}[t]
    \centering
    \includegraphics[width=\linewidth]{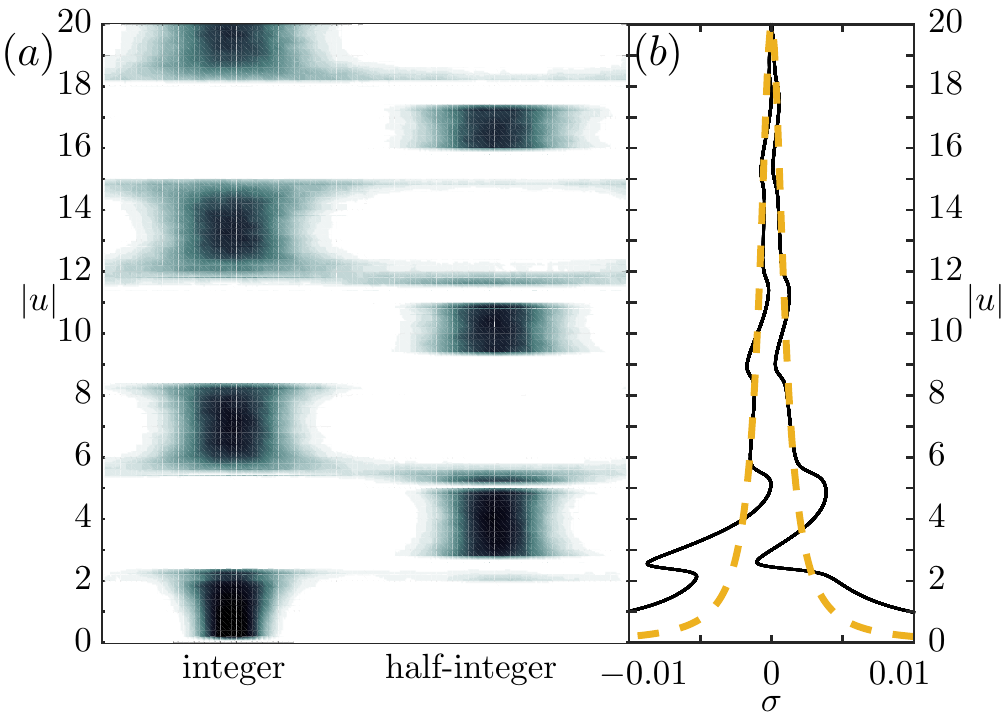}
    \caption{(a) Distribution of the SQUID phase $x$ (mapped to the interval $[-\tfrac14, \tfrac34]$) as a function of resonator amplitude $|u|$. (b) Theoretical response curve (solid black line) together with the unperturbed Lorentzian response (yellow dashed line). When $x$ is an integer, the resonator decouples from the SQUID, and the resonator response is very close to the Lorentzian. For half-integer $x$, the magnitude of the coupling to the SQUID is maximized, and the resonator response is highly non-linear. }
    \label{fig:switch}
\end{figure}

In order to further check the validity of the discussion in Sec.~\ref{sec:analytics}, we extracted the distribution of $x(\tau)$ for states stabilized at a certain envelope amplitude $|u|$. The result is shown in Fig.~\ref{fig:switch}.~(a), where switching between integer and half-integer $x_{\rm eq.}$ is clearly evident. The sections where no values are plotted are those $|u|$ where no stable state could be found, due to that $\partial|u|/\partial\sigma\rightarrow\infty$. For comparison, Fig.~\ref{fig:switch}~(b) includes the theoretical response curve together with the Lorentzian $f_0\gamma^{-1}\left(1+\sigma^2/\gamma^2\right)^{-1}$. In agreement with the discussion above, there is a clear correspondence between integer $x_{\rm eq.}$ and regions where the resonance curve is very close to the unperturbed response, whereas half-integer $x_{\rm eq.}$ coincide with highly nonlinear resonant response.

\begin{figure}
    \centering
    \includegraphics[width=\linewidth]{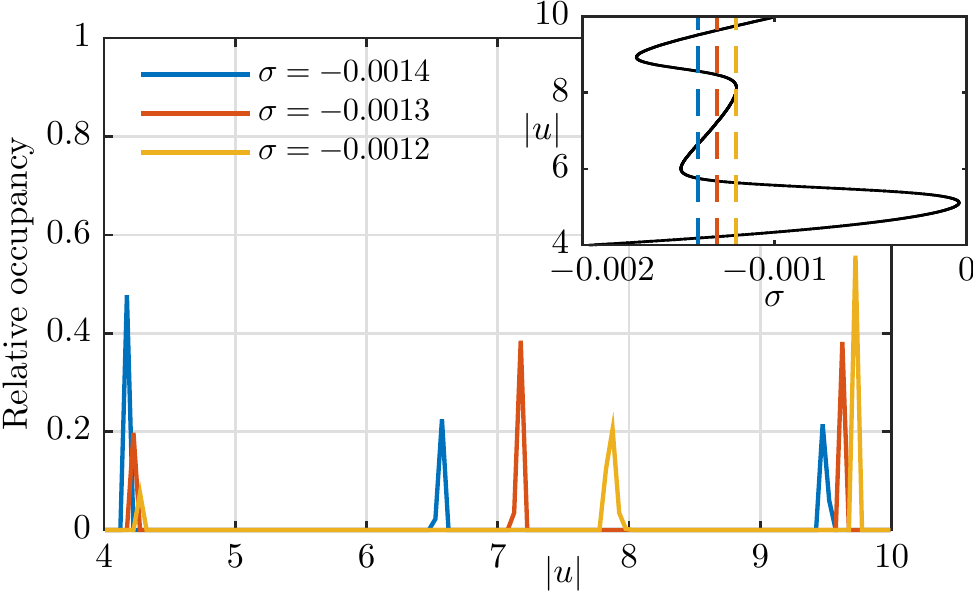}
    \caption{Distribution of resonator amplitudes for three values of the detuning $\sigma$. The inset shows a close-up of the relevant region of the analytical resonance curve, where the dashed lines indicate the values of $\sigma$ that were examined. For a given detuning and drive power, three different amplitudes can be observed.
    }
    \label{fig:multistab}
\end{figure}

Due to the presence of thermal noise, in Fig.~\ref{fig:resonance} expected hysteresis loops are smeared and there is very little difference between frequency sweeps up and down. Instead, the existence of multistability is proven by making a large number of measurements at the same detuning. To that end, several hundred trajectories were calculated, and the final resonator amplitude was recorded in each case. The initial state $(q(0),\dot q(0),x(0),\dot x(0))$ of the system was given by four random numbers, each uniformly distributed in the interval $(-10, 10)$. The result is shown in Fig.~\ref{fig:multistab}; the existence of multistability is clearly evident. Here, the detuning $\sigma$ was chosen to be one where the theoretical resonance curve indicates that several stable states might occur, see the inset of Fig.~\ref{fig:multistab}. 

\subsubsection{Zero-temperature limit}
\begin{figure}[t]
    \centering
    \includegraphics[width=\linewidth]{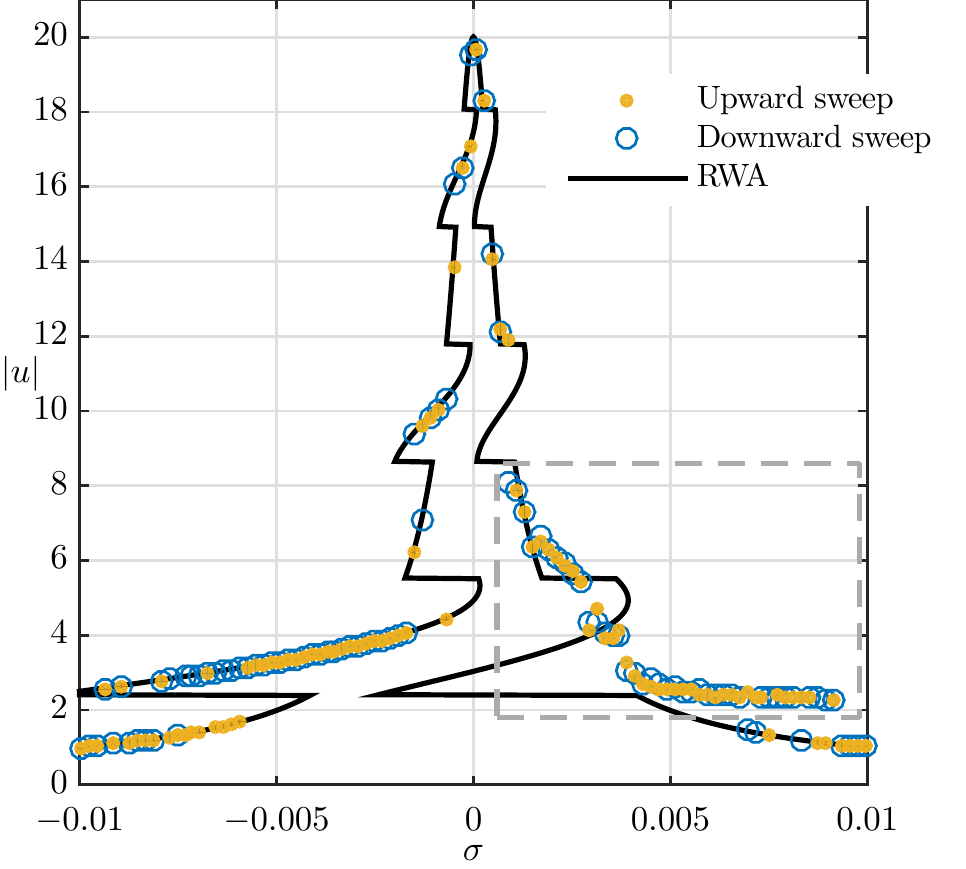}
    \caption{Resonant response at $T\rightarrow0$~K. The transitions between linear and nonlinear regimes are much sharper than in the case of finite temperatures. The dashed box indicates the anomalous region of detuning $\sigma$, where analytical and numerical results do not coincide.}
    \label{fig:T0}
\end{figure}

Finally, we consider the limit of millikelvin temperatures, such that $\mathcal D\simeq 10^{-6}$. With all other parameters as in Table~\ref{tab:params}, then $\alpha/\pi\eta\mathcal D=2.5\times10^4$. Consequently, the frequency shift $h(|u|)$ exhibits incredibly sharp features for $|u|$ such that $J_0(|u|)\approx 0$. The theoretical resonance curve inherits these sharp features, as can be seen in Fig.~\ref{fig:T0}. Still, for a large part of the response curve, the calculated response fits the theoretical curve surprisingly well. The exception is an anomalous region of positive $\sigma$, indicated in Fig.~\ref{fig:T0} by a dashed box. %In order to understand this anomaly, we study the time evolution of $q(\tau)$ and $x(\tau)$ for states in this region.

\begin{figure}[!t]
    \centering
    \includegraphics[width=\linewidth]{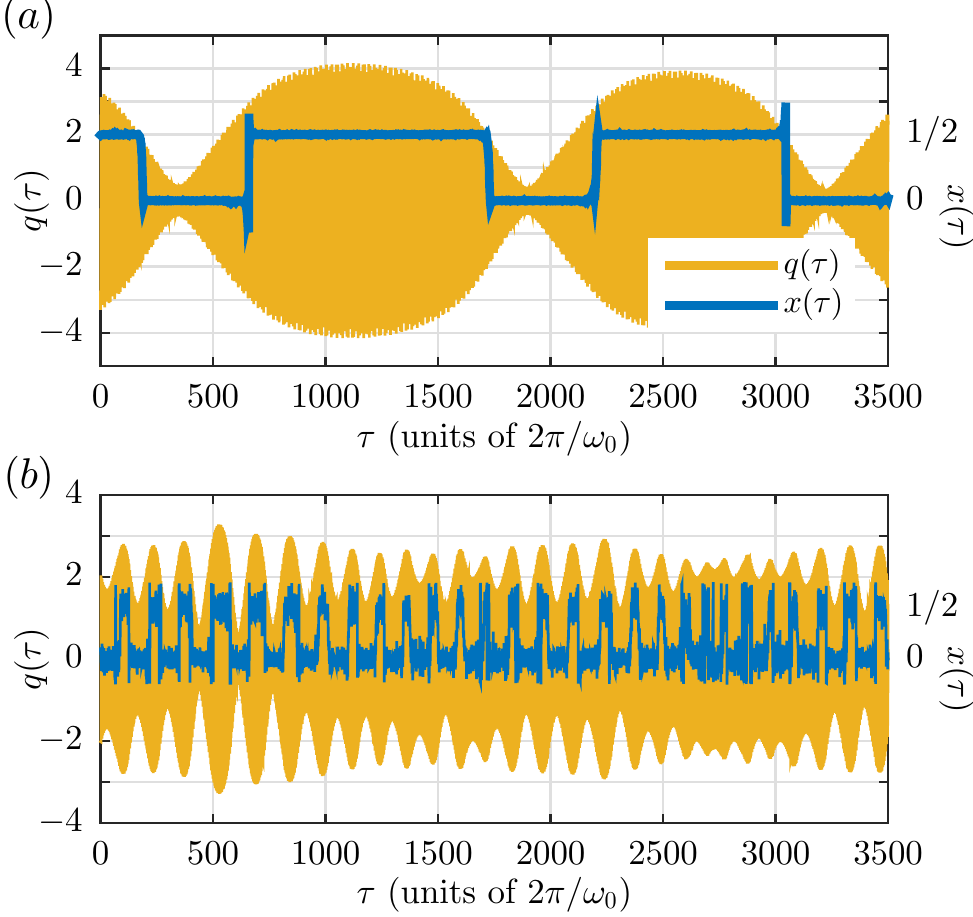}
    \caption{Time evolution for the LC-resonator amplitude $q$ and the SQUID flux $x$ (mapped to the interval $[-\tfrac14, \tfrac34]$), at $\sigma=0.003$ and (a) $T\rightarrow 0$~K (b) $T=3$~K. The response in the anomalous region is non-stationary, with resonator amplitude beats caused by $x$ switching between integer and half-integer values.}
    \label{fig:trajs}
\end{figure}

A typical time evolution of the oscillator coordinate $q$ and the flux particle position $x$ for detuning $\sigma=0.003$ are shown in Figure~\ref{fig:trajs}~(a). As can be seen, here in the anomalous part of the response, the system makes quasiperiodic transitions between integer and half-integer values of $x$, leading to beats in the resonator amplitude. The beats stem from the appearance of transient frequency components at $\epsilon h(|u|)$. These transients appear whenever a transition from integer to half-integer $x$ occurs, which causes the resonance to abruptly shift downwards. While the system remains at half-integer $x$ it is strongly nonlinear and can mix frequency components. Mixing with the drive at $\sigma\approx 1-\epsilon h(|u|)$  then causes a resultant which is on resonance, that consequently drives the oscillator at a shifted frequency, leading to the amplitude beats. 

Note that the corresponding phenomena cannot occur for negative detuning $\sigma<0$. Although the opposite process (half-integer to integer $x$) will lead to transients with positive frequency components, integer $x$ puts the resonator in the completely linear regime. Frequency mixing is then absent, and no component resurrecting the off-resonant motion can appear. Instead, only transient switching behavior is seen before the system reaches a stationary oscillatory state $\dot{u}=0$. 

This anomalous region of deviation between analytical and numerical results is seen also in Fig.~\ref{fig:resonance}, but only as less well-fitting data points near $\sigma=0.003$. In this case, the smoothing of $h(|u|)$ that is caused by the higher temperature makes the dynamics far less dramatic. While driving near $\sigma=0.003$ will still cause $q(\tau)$ to contain frequency components with negative detuning, thermal noise will smear the resulting amplitude beats, as seen in Fig.~\ref{fig:trajs}~(b), thus greatly decreasing the time the system spends in the nonlinear regime and limiting the frequency mixing. The higher noise level thus acts to \emph{stabilize} the resonator dynamics. This hints at the presence of stochastic resonance, in the broad sense of ``randomness that makes a nonlinearity less detrimental to a signal''~\cite{Mcdonnell_2009}.

\section{Outlook}
With superconducting circuit quantum electrodynamics being routinely done in the lab, the proposed system should be readily realized. Although the multistable response will only be visible for a particular range of drive powers, and the nonlinear parts of the resonance peak are very narrow, the current state of the art has matured to the point where detecting both these features is well within reach. A successful verification of the results in this Article would be the first experimental observation of induced nonlinearity in a diffusion-resonator system. Such an observation could stimulate further research into the influence of classical and quantum fluctuations in the interplay between harmonic oscillators and other dynamical systems.

\section{Acknowledgements}
We acknowledge helpful discussions with G\"oran Johansson and Jari Kinaret. This work was supported by the Swedish Research Council VR (AI), the Foundation for Strategic Research SSF (CR), and the Knut and Alice Wallenberg foundation (CR).

\end{document}